\documentclass[twocolumn,twocolappendix]{aastex63}

\usepackage{color}
\usepackage{amsmath}
\DeclareMathAlphabet\mathbfcal{OMS}{cmsy}{b}{n}
\usepackage{xspace}
\newcommand{\raptor}{{\tt RAPTOR}\xspace}
\newcommand{\mlody}{{\tt MLody}\xspace}
\newcommand{\symphony}{{\tt symphony}\xspace}

\received{12 Sep 2024}
\revised{22 Nov 2024}
\accepted{29 Nov 2024}
\submitjournal{ApJL}

\shorttitle{{\tt MLody}}
\shortauthors{Davelaar}

\graphicspath{{./}{}}

\begin{document}

\title{{\tt MLody} - Deep Learning Generated Polarized Synchrotron Coefficients}

\correspondingauthor{J. Davelaar}
\email{jdavelaar@princeton.edu}

\author[0000-0002-2685-2434]{J. Davelaar}
\affiliation{Department of Astrophysical Sciences, Peyton Hall, Princeton University, Princeton, NJ 08544, USA}
\affiliation{ NASA Hubble Fellowship Program, Einstein Fellow}
\affiliation{Center for Computational Astrophysics, Flatiron Institute, 162 Fifth Avenue, New York, NY 10010, USA}
\affiliation{Department of Astronomy and Columbia Astrophysics Laboratory, Columbia University, 550 W 120th St, New York, NY 10027, USA}

\begin{abstract}
Polarized synchrotron emission is a fundamental process in high-energy astrophysics, particularly in the environments around black holes and pulsars. Accurate modeling of this emission requires precise computation of the emission, absorption, rotation, and conversion coefficients, which are critical for radiative transfer simulations. Traditionally, these coefficients are derived using fit functions based on precomputed ground truth values. However, these fit functions often lack accuracy, particularly in specific plasma conditions not well represented in the datasets used to generate them. In this work, we introduce \mlody, a deep neural network designed to compute polarized synchrotron coefficients with high accuracy across a wide range of plasma parameters. We demonstrate {\tt MLody}’s capabilities by integrating it with a radiative transfer code to generate synthetic polarized synchrotron images for an accreting black hole simulation. Our results reveal significant differences, up to a factor of two, in both linear and circular polarization compared to traditional methods. These differences could have important implications for parameter estimation in Event Horizon Telescope observations, suggesting that \mlody could enhance the accuracy of future astrophysical analyses.
\end{abstract}


\section{Introduction}

Polarized synchrotron emission is a cornerstone of high-energy astrophysics, providing critical insights into some of the universe’s most extreme environments, such as the regions around black holes and pulsars. This type of emission is generated when charged particles undergo an acceleration perpendicular to their motion. High-energy astrophysical sources produce synchrotron emission via electrons gyrating around magnetic field lines. A key characteristic of synchrotron emission is its highly polarized nature \citep{rybicki1979}, making it a widely used tool for probing magnetic field structures and understanding the dynamics of these extreme astrophysical systems.

The Event Horizon Telescope (EHT) produced the first historic pictures of black holes, either in the center of Messier 87 \citep{eventhorizontelescopecollaboration2019} or from the center of our own galaxy, Sagittarius A* (SgrA*) \citep{eventhorizontelescope2022}. These observations, conducted at millimeter wavelengths, revealed polarized synchrotron emission originating from a relativistic population of electrons within the accretion flow surrounding the black hole. In the years following these historic results, the EHT also analyzed the polarization signatures coming from both Messier 87 \citep{eventhorizontelescopecollaboration2021,eventhorizontelescope2023} and SgrA* \citep{eventhorizontelescope2024}. These polarization studies provide critical insights into the magnetic field structures near event horizons, which are essential for understanding the dynamics of accretion flows and jet formation.

For the interpretation of the observations, the EHT uses general relativistic magnetohydrodynamical simulations of the accretion flow, which are post-processed by general relativistic ray tracing codes to compute synthetic images and spectra \citep{eventhorizontelescopecollaboration2019e,eventhorizontelescopecollaboration2022b,eventhorizontelescope2024b}. These radiative transfer codes generally all follow a similar algorithmic design \citep{Dexter2016,moscibrodzka2017,Bronzwaer2018,Bronzwaer2020}. In this approach, the code defines a virtual camera positioned outside of the simulations consisting of a predefined amount of pixels; each pixel is assigned a wave vector, which is used as the initial condition for solving the geodesic equation. The resulting trajectories (or rays) are then used to solve the radiation transfer equation 

\begin{equation}
\frac{{\rm d \bold{\mathbfcal{S}}}}{{\rm d}\lambda}   = 
         \begin{pmatrix}
           j_I \\
           j_Q \\
           j_U \\
           j_V
         \end{pmatrix} -
         \begin{pmatrix}
           \alpha_I & \alpha_Q & \alpha_U & \alpha_V \\
           \alpha_Q & \alpha_I & \rho_V   & -\rho_U  \\
           \alpha_U & -\rho_V  & \alpha_I & \rho_Q   \\
           \alpha_V & \rho_U   & -\rho_Q  & \alpha_I
         \end{pmatrix}
         \begin{pmatrix}
           \mathcal{I} \\
           \mathcal{Q} \\
           \mathcal{U} \\
           \mathcal{V}
         \end{pmatrix}.
\label{eqn:plasma_interaction}
\end{equation}

The left hand side of Eqn. \ref{eqn:plasma_interaction}, contains; $\lambda$ the affine parameter, and $\mathbfcal{S}$ the Lorentz invariant Stokes vector given by

\begin{equation}
\mathbfcal{S} = \frac{\bold{S}_{\nu}}{\nu^3}  = \frac{1}{\nu^3}    
 \begin{pmatrix}
           {I_{\nu}} \\
           {Q_{\nu}} \\
           {U_{\nu}} \\
           {V_{\nu}}
         \end{pmatrix}
\end{equation}
         
here $S_{\nu}$ is the specific intensity for every Stokes component $I,Q,U,V$, and $\nu$ is the frequency. The right-hand side of Eqn. \ref{eqn:plasma_interaction} depends on the polarized synchrotron coefficients, first the emission coefficients $j_I, j_Q, j_U, j_V$, secondly absorption coefficients $\alpha_I, \alpha_Q, \alpha_U, \alpha_V$, and finally the rotation and conversion coefficients $\rho_Q, \rho_U, \rho_V$. Polarized radiative-transfer computations depend on the orientation of the frame in which the Stokes vector can be defined. This requires constructing a frame at the observer's location (the camera) and any location where the ray interacts with radiating matter. A suitable choice for the latter is one where Stokes $U$ is aligned with the magnetic field vector, resulting in $j_U=\alpha_U=\rho_U = 0$, and $j_Q < 0$.

At every time step along the ray, the coefficients are computed based on the local plasma conditions. Calculating these coefficients involves an integral over momentum space and a sum over the synchrotron harmonics, see, e.g., \cite{leung2011,Dexter2016}. The integral over momentum space depends on the shape of the electron distribution function, which gives the particle number at a given energy (or momentum). Common choices for the electron distribution function include the relativistic Maxwellian, $\kappa$-distribution functions, and power-law distribution functions. Computing this integral accurately over the full range of possible plasma conditions is computationally very demanding. Therefore, many forms of approximate ``fit formula" are used instead of performing the full complex computation for every step. Common fit formulas in the literature that are tested against the exact numerical solution are \cite{Shcherbakov2008,leung2011,Dexter2016,pandya2016,Marszewski2021}. The weakness of this approach is that fit functions are by construction approximate and are only constructed using a limited range of the parameter domain, e.g., \cite{pandya2016}\footnote{We will compare our results mainly to the work by \cite{pandya2016} since it is the most commonly used fit formula within the Event Horizon Telescope Collaboration.} only considered dimensionless electron temperatures between 2 and 40, which limits their accuracy in regions where absorption and Faraday rotation dominate at lower temperatures. Additionally in the ultra-relativistic synchrotron limit the integral over the distribution function can be analytically approximated, e.g. \cite{mahadevan1996,Dexter2016}.

To address the limitations of traditional fit functions, we developed \mlody, a deep neural network designed to enhance the accuracy of polarized synchrotron coefficient computations across a diverse range of plasma conditions, extending high accuracy to the lower temperature regime. The code is publicly availbe on {\tt GitHub}\footnote{https://github.com/jordydavelaar/mlody}. In Section \ref{sec:methods}, we will outline the training data generation, the network architecture, and how we interfaced \mlody with \raptor\footnote{Publicly available at https://github.com/jordydavelaar/raptor} \citep{Bronzwaer2018,Bronzwaer2020}. In Section \ref{sec:results}, we will show the network's training results, and present a use case of \mlody's capabilities by computing synthetic images of black hole accretion flow simulations. In Section \ref{sec:conclusion} we will draw our conclusions.

\section{Methods}\label{sec:methods}

\subsection{Training Data Generation}

We used the publicly available code \symphony to generate a large sample of polarized synchrotron radiation coefficients. For a detailed explanation of \symphony, see \cite{pandya2016} and references therein. The coefficients generally depend on four variables: the electron number density $n_{\rm e}$, the frequency $\nu$, the magnetic field strength, $B$, and the angle between the wave vector $k$ and magnetic field, referred to as the pitch angle, $\theta_{\rm B}$. Additionally, certain variables are connected to the specific distribution function. For a thermal Maxwellian, this includes the dimensionless electron temperature $\Theta_{\rm e}=\frac{kT}{m_{\rm e} c^2}$, with $k$ the Boltzmann constant, $m_{\rm e}$ the electron mass, and $c$ the speed of light. 

Naively this would require a five-dimensional data cube for our input vector. However, by utilizing the exact form of the integrals solved by \symphony, we can reduce this to three dimensions. First, all coefficients depend linearly on electron number density; therefore, we can always multiply with density after computing the coefficients. Secondly, we can define  $X_{\nu} = \nu / \nu_{\rm c}$, where $\nu_{\rm c} = \frac{eB}{2\pi m c}$ is the electron cyclotron frequency, this combines the frequency and magnetic field, resulting in another reduction of our input dimension. The coefficients can then be written in a universal form 

\begin{eqnarray}\label{eqn:units}
    j_{S} =  \frac{n_{\rm e} e^2 \nu}{c} J_{S}( X, \theta_B, \theta_e)\\
    \alpha_{S} =  \frac{n_{\rm e} e^2}{\nu_{\rm c} m_{\rm e} c} A_{S}( X_{\nu}, \theta_{\rm B}, \Theta_{\rm e})\\
    \rho_{S} = n_e R_{S}( X, \theta_B, \theta_e), \label{eqn:units1}
\end{eqnarray}
where $J_{S}$ and $A_S$ are dimensionless functions. For training data of our network we will computed $J_{S}$, $A_{S}$ and $R_{S}$ by sampling the three input variables randomly over the following ranges $X_\nu \in [1,10^6]$, {$\theta_{\rm B} \in [0,2 \pi]$}, and $\Theta_{\rm e} \in [10^{-2},10^{2}]$. In total, the training data includes nearly 20 million coefficients.  

\subsection{Network Architecture and Learning Strategy}

To build our neural network we use {\tt Keras} \citep{keras2015} with a {\tt TensorFlow} \citep{tensorflow2015} backend. The network architecture is a fully connected deep neural network with nine Dense layers. These layers are divided into three groups, each with 100, 200, and 300 neurons, respectively. For every Dense layer, we use a rectified linear unit (ReLU) activation function, described by $f(x) = \max(0, x)$ \citep{nair2010}. ReLU activation functions are essential for introducing non-linearities into the network, enabling it to learn complex underlying functions. After the last group of Dense Layers, we have our output layer which consists of eight neurons and utilizes a linear activation function. For a summary of our network architecture see Table \ref{tab:network}.

\begin{table}[]
    \centering
    \begin{tabular}{c|c|c}
         Layer & Neurons & Activation  \\
         \hline
         Input   & 3   & -    \\
         Dense 1-3 & 100  & ReLU \\
         Dense 4-6 & 200 & ReLU \\
         Dense 7-9 & 300 & ReLU \\
         Output  & 8   & linear 
    \end{tabular}
    \caption{Summary of the network architecture. Listed are the various layers used, the number of neurons per layer, and the activation functions.}
    \label{tab:network}
\end{table}

We split the data 80/20 between training and validation. Additionally, we keep 20,000 randomly selected samples aside for testing after training is finished. The validation set is used to monitor for overfitting. Overfitting occurs when the network perfectly replicates the training data without learning how to generalize, which would result in poor performance on unseen data. By comparing the loss on the training set with the validation set, we ensure the network is learning the true underlying correlations rather than simply memorizing the training data. If the training loss decreases while the validation loss increases, this indicates overfitting.

We employed the Adam optimizer \citep{kingma2014}, and the network uses an exponential-decay learning-rate schedule with an initial learning rate of $10^{-4}$, a decay rate of 0.95, and a decay step of 10,000. This means the learning rate decreases by a factor of 0.95 for every 10,000 gradient descent steps. The loss function used was mean squared error:

\begin{equation}
    L = \frac{1}{N} \Sigma^N_{\rm i} |y_{\rm i} - \hat{y_{\rm i}}|^2,
\end{equation}
where $y_{\rm i}$ is the true value of the $i$'th sample, $\hat{y_{\rm i}}$ is corresponding network prediction, and $N$ is the total number of data points. The network was trained for 1,000 epochs and used a batch size of 2,000. 

Before training, the input vector $X$ and output $\hat{y}$ are transformed and rescaled so that their values map between $-1$ and $1$ and their mean is zero. In the case of $X_\nu$, $\Theta_e$, $J_S$, $A_S$, we 1) take the logarithm, 2) recenter them to have a mean of zero, and 3) normalize with their maximum value. For $\theta_{\rm B}$, we train on $\cos \theta_{\rm B}$, which is naturally normalized between $[-1,1]$.  Since $\rho_Q$ and $\rho_V$, both have positive as well as negative values, we did not use a logarithmic function; instead, their scaling was performed by first applying $X = {\rm sign}(\rho_S) | \rho_S |^{0.25} $, then recentering and normalizing by the maximum value.

\subsection{Integration with \raptor}

After the network was successfully trained, we interfaced \mlody with \raptor by using the publicly available {\tt keras2cpp} \footnote{See https://github.com/gosha20777/keras2cpp} library. This library is a {\tt C++} framework that reads in a model file that contains all the information on the network architecture and the neuron weights. The library also provides functions that can be called by \raptor to compute new outputs. For this final step, the code converts the plasma quantities to the input vector format of \mlody, obtains an output vector, and computes $J_{S}$, $A_{S}$ and $R_{S}$ by rescaling the output and add units via Equations \ref{eqn:units}-\ref{eqn:units1}. In case the input variables for \mlody are outside the range of the training data, we default to using the fit functions.

\section{Results}\label{sec:results}

\subsection{Network Training and Performance Assessment}

\begin{figure}
    \centering
    \includegraphics[width=0.5\textwidth]{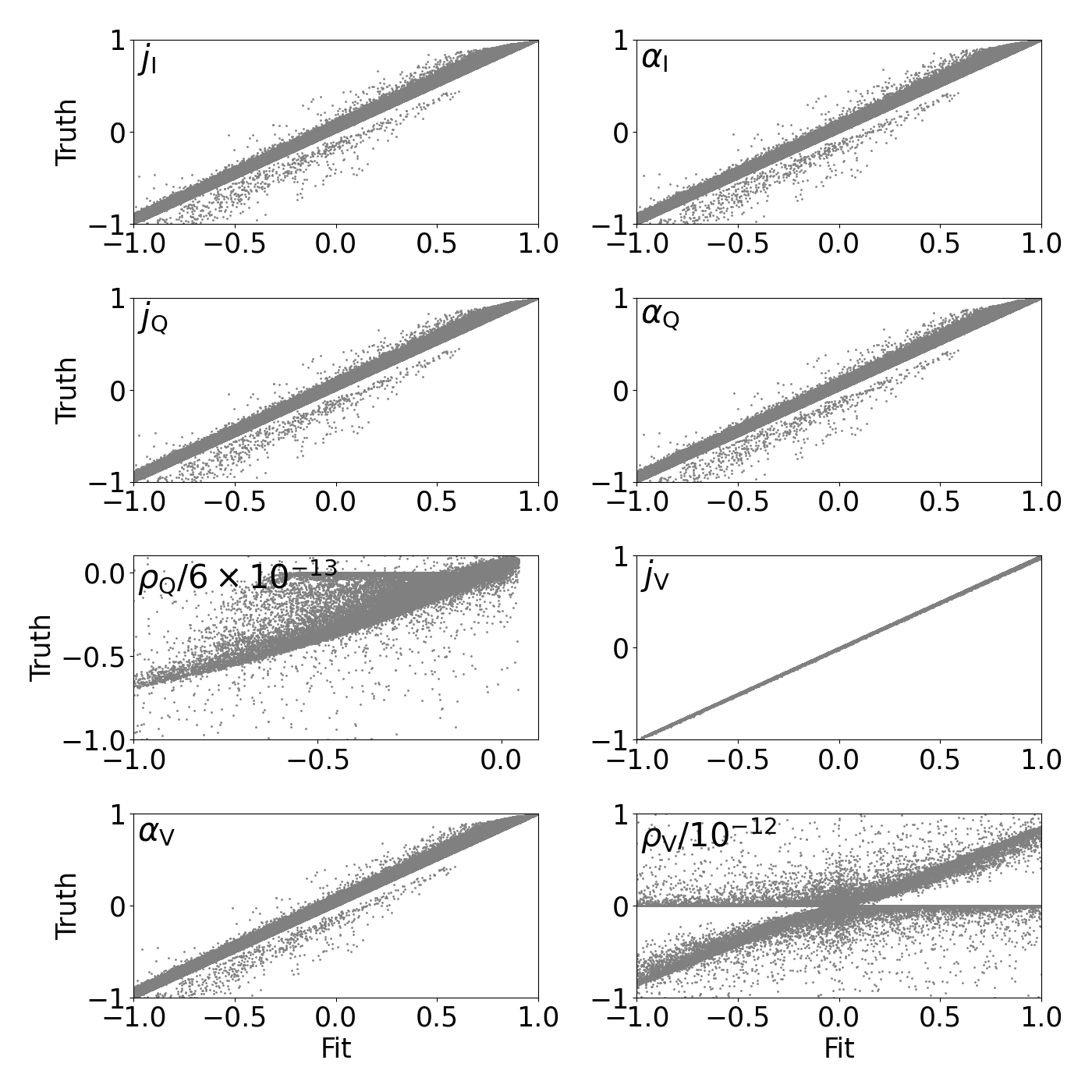}
    \caption{The ground truth values of \mlody are plotted as a function of their corresponding fit function values. A perfect diagonal would mean identical values for both methods. The largest discrepancies are found for the rotation and conversion coefficients, $\rho_Q$ and $\rho_V$. }
    \label{fig:fit-acc}
\end{figure}

The training data set, computed with \symphony, consists of nearly 20 million coefficients. To assess the performance of the training data against the fit formula from \cite{pandya2016}, we compared the results using both approaches, as illustrated in Fig. \ref{fig:fit-acc}. In an ideal scenario, all values would align along the diagonal, indicating perfect agreement between the fit function on the $x$ axis and the \symphony output on the $y$ axis. The emission and absorption coefficients all show a wider band around the optimal diagonal, with the largest discrepancies at low $X_{\nu}$ and electron temperatures, which are outside the data range on which the fit functions were constructed. Moreover, the rotation and conversion coefficients display significant deviations across the entire parameter range, including incorrect signs.  The deviations are found to be largest for small electron temperatures $\Theta_{\rm e} < 1$ and $\Theta_{\rm e}^2 / X_{\rm nu} < 1$, which indicates that this is at lower electron energies. These deviations with fit functions arise since in the literature they are constructed by either computing them analytically in the high-energy limit \citep{Shcherbakov2008,pandya2016,Marszewski2021} In this case the lower energy limit is poorly incorporated and the fit functions need to extrapolation outside of their range where the high-energy assumption holds.

\begin{figure}
    \centering
    \includegraphics[width=\columnwidth]{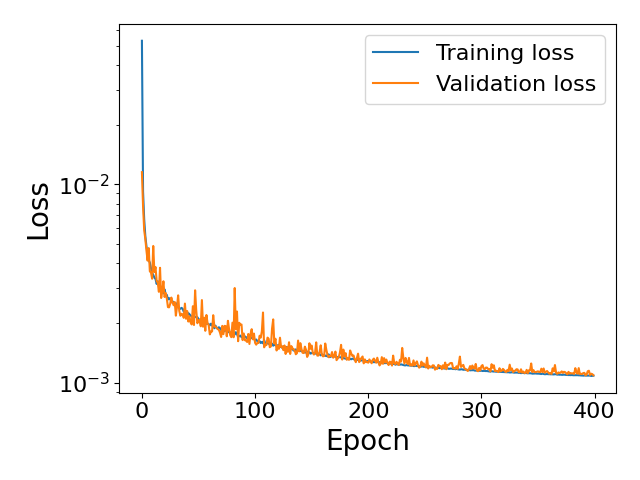}
    \caption{Loss as a function of epoch for the training and validation set. Both sets perform equally well, which is a clear indicator that we do not overfit the training data.}
    \label{fig:loss}
\end{figure}

Next, we train the network for 1,000 epochs and evaluate its performance. Fig. \ref{fig:loss} shows the training and validation Loss as a function epoch. Both losses steadily decrease with epoch and follow similar trajectories, indicating good network performance and a lack of overfitting. At the end of the training, the Loss reaches a value of $5 \times10^{-4}$. We additionally also compute the mean absolute relative error of the network, and for both the training as well as the validation set, this is below one percent at the end of training.

\begin{figure}
    \centering
    \includegraphics[width=0.5\textwidth]{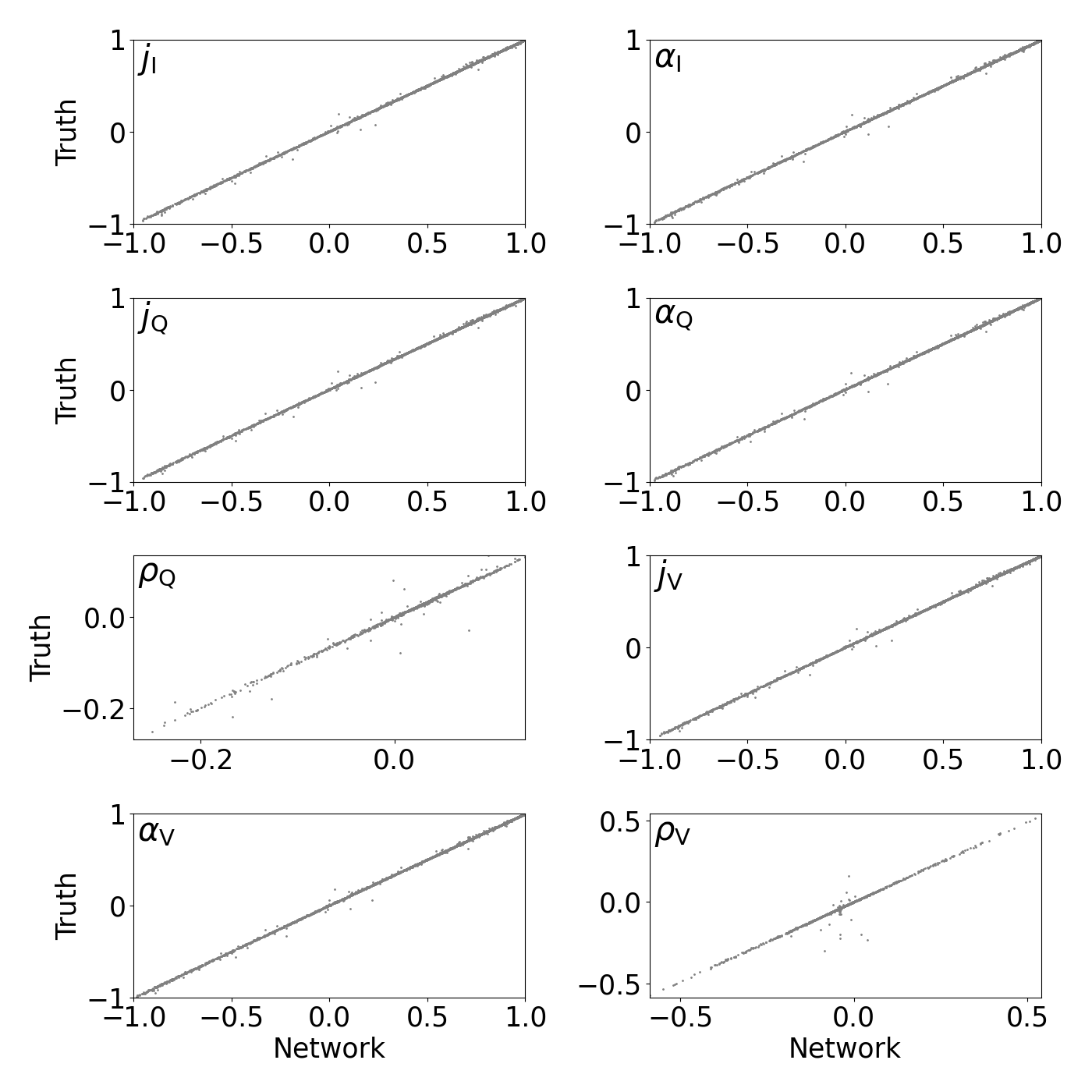}
    \caption{The same as Fig. \ref{fig:fit-acc} but now as a function of the predictions of \mlody. Compared to Fig. \ref{fig:fit-acc} the performance is much higher, especially for the conversion and rotation coefficients.}
    \label{fig:net-acc}
\end{figure}

To further evaluate \mlody’s performance, we compared the network’s predictions to the ground truth from \symphony in Fig. \ref{fig:net-acc} by using the 20,000 samples in the test data set. Compared to Fig. \ref{fig:fit-acc}, \mlody outperforms the traditional fit functions across all coefficients.  The most notable improvements can be seen in the rotation and conversion coefficients $\rho_Q$ and $\rho_V$.

\subsection{Performance Comparison for Black Hole Accretion Modeling}

As an example of \mlody's capabilities, we integrated the network with \raptor to compute synthetic images of a black hole accretion flow simulation. We use a simulation snapshot run with the GRMHD code {\tt BHAC} \citep{porth2017}. The black hole simulation is in the Magnetically Arrested Disk (MAD) regime \cite{narayan2003,tchekhovskoy2011}; see \cite{davelaar2023} for simulation details. Meaning that the black hole reaches a saturated state in the amount of magnetic flux on the horizon. The spin of the black hole is $a=0.9375$. Ideal GRMHD simulations are scale-free and typically do not evolve the electron thermodynamics explicitly. Therefore, plasma variables must be rescaled to achieve a physical scale in post-processing. For this test, we chose scaling parameters to model the black hole at the center of our galaxy, Sagittarius A*. The black hole mass and distance are constrained by long-term observations of stellar orbits in the galactic nucleus, $M_{\rm BH} = 4.14\times10^6 M_\odot$, $d=8.127$ kpc \citep{do2019,gravity2019}. The mass accretion rate we set to $\dot{M} = 10^{-8} M_\odot {\rm yr}^{-1}$, resulting $\approx1$ Jy of flux at 230 GHz. 

Finally, the electron temperature was parameterized using the prescription from \cite{moscibrodzka2016}
which relates the electron temperature $T_{\rm e}$ to the proton temperatures $T_{\rm p}$ via

\begin{subequations} \label{eq-temp-rat}
\begin{align}
T_{\rm ratio} &= \frac{1}{1+\beta_{\rm p}^2} + R \frac{\beta_{\rm p}^2}{1+\beta_{\rm p}^2},\\
\Theta_{\rm e} &= k_{\rm b} T_{\rm e}/m_{\rm e} c^2 =  \frac{U(\gamma_{\rm ad} - 1) m_{\rm p}/m_{\rm e}}{\rho \left(1 + T_{\rm ratio} \right)},
\end{align}
\end{subequations}
where $m_{\rm p}$ the proton mass, $m_{\rm e}$ the electron mass, $U$ the internal energy, and $\Theta_{\rm e}$ the dimensionless electron temperature. The free parameter $R$ determines the temperature ratio in regions where $\beta_{\rm p}= 2P/B^2 \gg 1$. The choice of $R$ will change the morphology of the resulting image, e.g., the disk dominates the emission for $R=1$ while the jet dominates if $R\gg 1$. For our test case, we will set $R=40$, meaning a jet-dominated model. Finally, we set the inclination of the observer relative to the black hole spin axis at 30 degrees. We then generate synthetic polarized spectra and images between $10^{10}- 10^{14}$ Hz using both \mlody and the fit function from \cite{pandya2016}.

\begin{figure}
    \centering
    \includegraphics[width=\columnwidth]{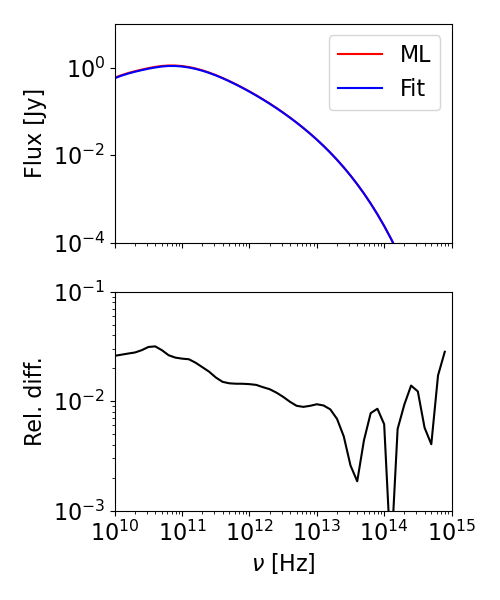}
    \caption{Spectrum of Stokes $I$ for both the \mlody generated coefficients as well as the fit formula (top panel). The bottom panel shows the relative difference between the two spectra, |Fit-ML|/Fit. The overall difference is small, and nearly independent of frequency.}
    \label{fig:spec_I}
\end{figure}

We computed spectra for all Stokes variables and compared them to results obtained using the fit function method. In Fig. \ref{fig:spec_I} we show the spectra for Stokes $I$. Overall, we found good agreement between both methods, with relative differences remaining close to one percent across the entire frequency range, as shown in the bottom panel. This small discrepancy is not unexpected given that there is a relatively small difference between the ground truth and the fit formula for Stokes $I$ (see Fig. \ref{fig:fit-acc}) in contrast to the rotation and conversion coefficients which only affect Stokes $Q$, $U$ and $V$.

Given the poorer performance of the fit function for the rotation and conversion coefficients, we expect larger discrepancies in the polarized quantities. We computed both image-integrated and pixel-integrated values of linear and circular polarization. The difference lies in when the summation over pixel values occurs: we either sum Stokes $Q$ and $U$ first and then compute the total linear polarization $\sqrt{Q^2 + U^2}$, or we compute $\sqrt{Q^2 + U^2}$ per pixel and then the sum. The former approach mimics an under-resolved measurement, such as one from a single telescope with a large field of view, while the latter represents a perfectly resolved measurement. 

\begin{figure}
    \centering
    \includegraphics[width=\columnwidth]{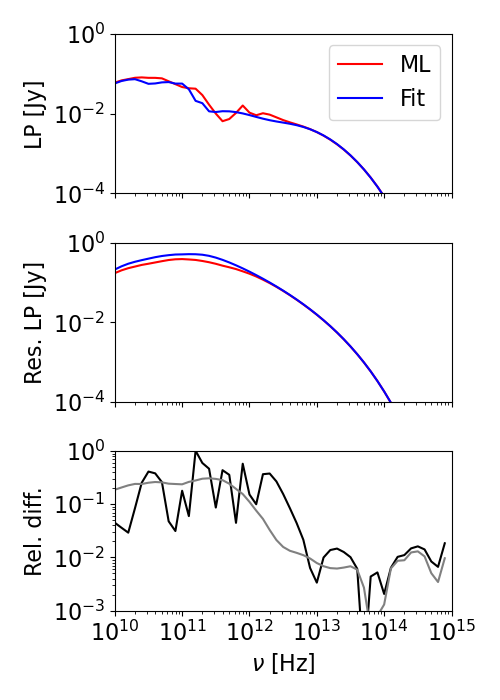}
    \caption{Same as Fig. \ref{fig:spec_I} but now for linear polarization, $\sqrt{S_Q^2 + S_U^2}$. Here, we find larger relative differences compared to Stokes $I$, up to 30\%, peaking at $10^{11}$ Hz, and 1\% at higher frequencies.}
    \label{fig:spec_LP}
\end{figure}

Fig. \ref{fig:spec_LP} shows the linear polarization $\sqrt{Q^2 + U^2}$ for both unresolved (top panel) and resolved (middle panel) cases, with the relative difference between the two methods shown in the bottom panel. In both cases, we find maximum differences of $~10$\%, where \mlody generates less linear polarization than the fit functions. Since the sign of $\rho_Q$ correlates with the Faraday Rotation Measure we expect deviations between \mlody and the fit functions. The rotation measure is given by $RM= \frac{\chi_1 - \chi_2}{\lambda_1^2 - \lambda_2^2}$, with $\chi=0.5 \arctan(Q/U)$ the EVPA and $\lambda$ the wavelength. We use the image integrated $S_Q$ and $S_U$ at $100$ GHz and $158$ GHz to compute the $RM$. We find slightly lower values for the $RM$ but with the same sign for \mlody compared to the fit functions, namely,  $RM=3.7\times10^4$ rad/m$^2$ and $RM=2.5\times10^4$ rad/m$^2$, respectively.

\begin{figure}
    \centering
    \includegraphics[width=\columnwidth]{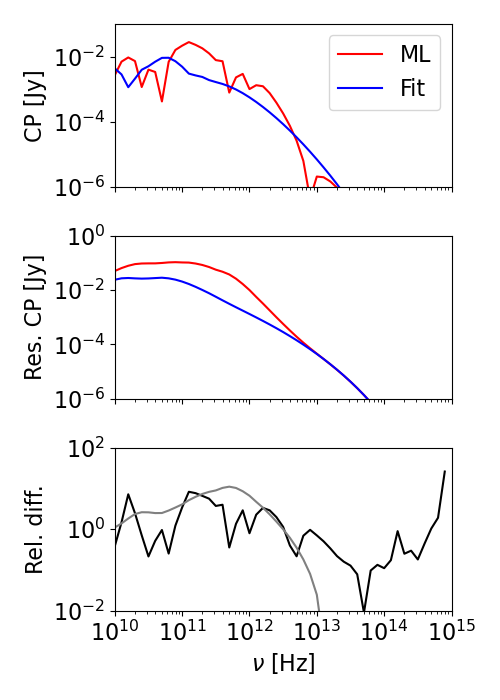}
    \caption{Same as Fig. \ref{fig:spec_I} but now for circular polarization, $|S_V^2|$. Compared to linear polarization the relative difference is comparable in the majority of the spectrum for CP, of the order of 10-100\%. for the resolved CP, the relative difference is only large in the low-frequency part of the spectrum and drops after $10^{12}$ Hz.}
    \label{fig:spec_CP}
\end{figure}

Fig. \ref{fig:spec_CP} presents the circular polarization results, $|V|$. The top panel shows the unresolved circular polarization, the middle panel the resolved circular polarization, and the bottom panel the relative difference between the two models. The relative difference with the fit function is largest for Stokes $V$, reaching up to 100\% for most frequencies, with \mlody generating more circular polarization than the fit functions.
 
\begin{figure*}
    \centering
    \includegraphics[width=0.9\textwidth]{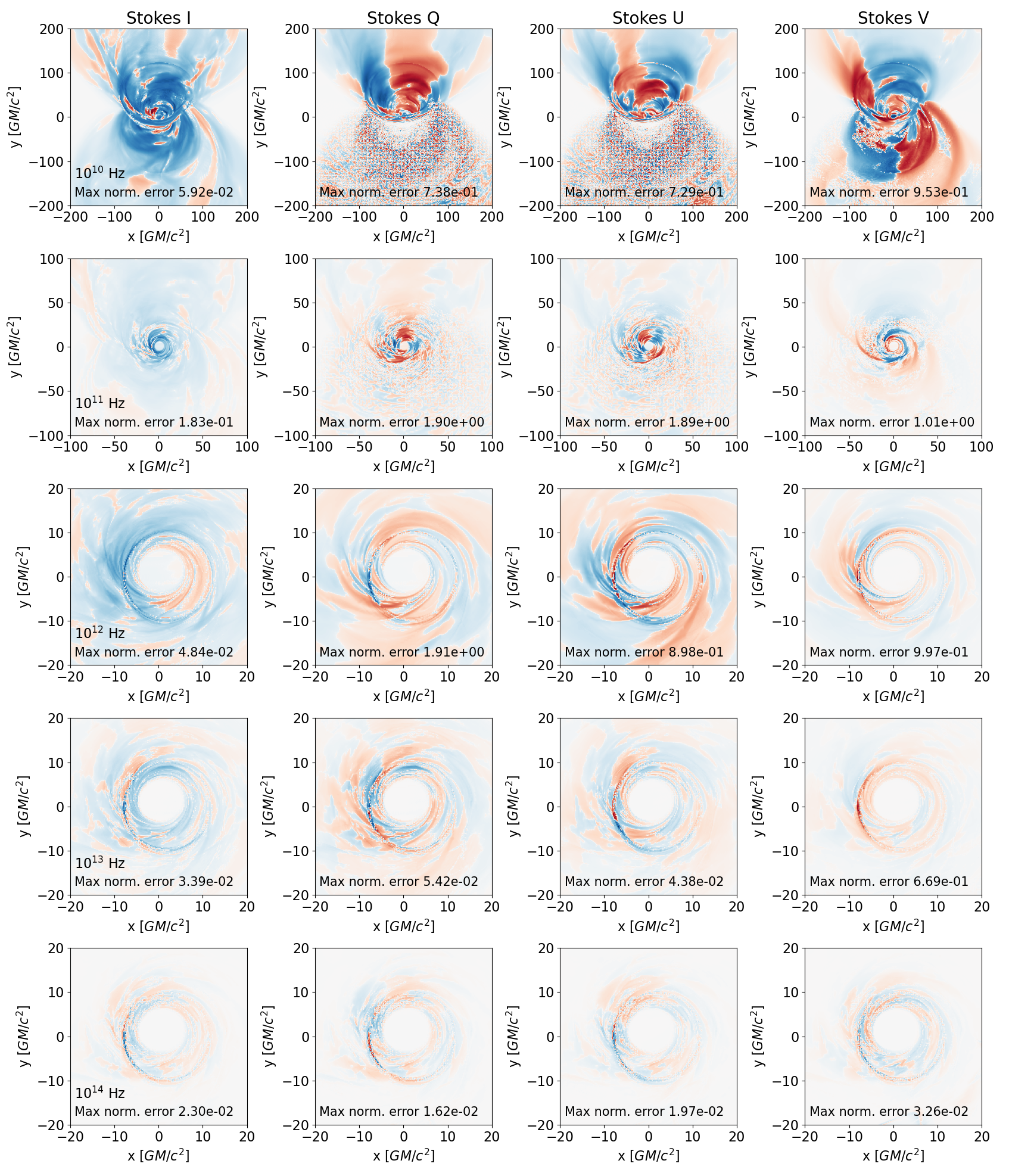}
    \caption{Pixel-to-pixel difference maps on synthetic polarized synchrotron images either computed with \mlody or fit functions. The maps show sign(Fit-\mlody)$|$Fit-\mlody$|^{0.25}$, where red are positive error values and blue negative error values, all images are normalized to the maximum absolute pixel value. The first, second, third, and fourth columns are Stokes $I$, $Q$, $U$, and $V$, respectively. Rows correspond to difference frequencies, top to bottom; $\nu = 10^{10}, 10^{11}, 10^{12}, 10^{13}, 10^{14}$ Hz. Within each panel we show the maximum normalized error given by max($|$Fit-ML$|$)/max($|$ML$|$). The largest differences are found for Stokes $Q$, $U$, and $V$ at frequencies$\nu=10^{11}-10^{12}$ Hz.}
    \label{fig:images}
\end{figure*}

In Fig.\ref{fig:images}, we show the pixel-to-pixel residuals between synthetic images computed with and without \mlody at $\nu=10^{10},10^{11},10^{12},10^{13},10^{14}$ Hz. Each image is normalized to its largest pixel value, with an inset showing the absolute error divided by the maximum pixel value of the image. The first column shows residual images for Stokes $I$. As expected from the spectra, the residuals are small, generally below $~0.1$.  For Stokes $Q$, $U$, and $V$, larger differences are observed. This is in agreement with the larger relative differences found in the linear and circular polarization spectra, see Figs. \ref{fig:spec_LP} and \ref{fig:spec_CP}.  The maximum differences are found around $10^{12}$ Hz and decrease at higher frequencies. The largest discrepancies are seen either far out in the jet, see $10^{10}$ Hz, where the temperature and $X_\nu$ decrease, or close to the black hole for higher frequencies. At the higher frequencies, $10^{13}$ Hz and $10^{14}$ Hz, we find a flip in the sign of the offset in stokes V on the photon ring, however since the offset is small, the sign found in the individual images is the same for both \mlody and the fit functions.

The computational speed of \raptor is affected by the current implementation of \mlody, making the code a factor of a few slower compared to when it is run with the fit functions. However, the keras2cpp library used to connect \mlody with \raptor is not yet optimized for our specific use case. Improving the computational performance of the network is ongoing, and speed-ups are expected to be needed to make the network more efficient for EHT analysis. However, improvements observed with \mlody suggest that deep learning techniques could become integral to modeling efforts for the study of black hole environments.

\section{Conclusion}\label{sec:conclusion}

In this letter, we introduced \mlody, a novel deep neural network specifically designed to generate polarized synchrotron coefficients with high accuracy, particularly tailored for black hole accretion studies. Trained on millions of coefficients from the publicly available {\tt symphony} code, \mlody outperforms traditional fit functions across a wide range of plasma conditions. Our results not only demonstrate the increased performance of \mlody over traditional methods but also highlight its potential to increase the accuracy of the analysis of future EHT observations and related astrophysical research.

This work focused on the thermal Maxwellian electron distribution function, demonstrating significant improvements in the accuracy of polarization modeling. Future developments will extend \mlody to support $\kappa$ and power-law distribution functions, as well as other more complex electron distributions, such as anisotropic distribution function \citep{galishnikova2023} or distribution function computed by first-principle kinetic plasma simulations. These expansions will further enhance \mlody’s applicability to a broader range of astrophysical scenarios.

By interfacing \mlody with the radiative transfer code \raptor, we generated synthetic images from a black hole accretion simulation and observed discrepancies in polarization spectra—ranging from 10\% to 100\%—compared to results derived from fit functions. These differences are substantial and could critically impact the parameter estimations used in EHT observations, particularly for M87 and SgrA* \citep{eventhorizontelescopecollaboration2021,eventhorizontelescope2023,eventhorizontelescope2024}, where models are often validated against specific polarization thresholds. While a comprehensive comparison with EHT data is beyond the scope of this letter, our findings suggest that incorporating \mlody could lead to more accurate model-to-data comparisons and potentially refine the interpretation of EHT results.

In conclusion, \mlody demonstrates enhanced accuracy in modeling polarized synchrotron emission, providing a more precise tool for studying black hole environments. Its application in upcoming observational campaigns, including those by the next-generation Event Horizon Telescope, may lead to more accurate interpretations of black hole observations.

\acknowledgments

J.D. thanks the anonymous referee for their helpful and constructive comments on the manuscript. J.D. is supported by a NASA Hubble Fellowship Program Einstein Fellowship. J.D. acknowledges support from a Joint Columbia University and Flatiron Institute Postdoctoral Fellowship. J.D. thanks Luc Hendriks for fruitful discussions and suggestions on machine learning networks. J.D. acknowledges the work by Georgy Perevozchikov, GitHub user gosha20777, for their public {\tt keras2cpp} library. The Simons Foundation supports research at the Flatiron Institute. The computational resources in this work were provided by facilities supported by the Scientific Computing Core at the Flatiron Institute, a division of the Simons Foundation.

\software{ {\tt RAPTOR} \citep{Bronzwaer2018,Bronzwaer2020}}, {\tt Keras} \citep{keras2015}, {\tt TensorFlow} \citep{tensorflow2015}, \symphony \citep{pandya2016}, {\tt BHAC} \citep{porth2017}, {\tt python} \citep{oliphant2007,millman2011}, {\tt scipy} \citep{jones2001}, {\tt numpy} \citep{vanderwalt2011}, {\tt matplotlib} \citep{hunter2007},

\end{document}